\documentclass[notitlepage,twocolumn,prl,tightenlines,nofootinbib,superscriptaddress,showpacs]{revtex4}

\usepackage{amsmath}
\usepackage{amssymb,amsfonts,latexsym,MnSymbol}
\usepackage{bm}
\usepackage[mathcal]{euscript}
\usepackage{graphicx}
\usepackage{epsfig}
\usepackage{color}

\newcommand{\ie}{\textit{i.e. }}

\begin{document}

\title{Creep motion of an intruder within a granular glass close to jamming}

\author{R. Candelier}
\affiliation{SPEC, CEA-Saclay, URA 2464 CNRS, 91 191 Gif-sur-Yvette, France}
\author{O. Dauchot}
\affiliation{SPEC, CEA-Saclay, URA 2464 CNRS, 91 191 Gif-sur-Yvette, France}

\begin{abstract}
We investigate experimentally the dynamics of an intruder dragged by a constant force in an assembly of horizontally vibrated grains close to jamming. At moderate packing fractions, the intruder moves rapidly as soon as the force is applied. Above some threshold value of the packing fraction which increases with the applied force, the intruder exhibits an intermittent creep motion with strong fluctuations reminiscent of a "crackling noise" signal. These fluctuations behave in a critical manner at the jamming transition $\phi_J$ unveilled in a previous study~\cite{lechenault2008csa}. The transition separates a regime where the intruder motion is dominated by local free volume rearrangements from a regime where the instantaneous displacement field is strongly heterogeneous and resemble the force chains patterns observed in dense granular packings.
\end{abstract}

\maketitle

The understanding of mechanical properties of amorphous media such as granular media, foams, emulsions, suspensions or structural glasses raised a formidable interest in the past decades~\cite{jaeger1996gsl}. At high packing fractions, such materials eventually jam and sustain a finite shear stress before yielding~\cite{dollet2005tdf,habdas2004fmp,xu2006mys}. Several experiments on granular systems have emphasized the role of dynamical heterogeneities in the glassy like increase of the structural relaxation time~\cite{marty2005sac}. More recently it was shown that dynamical heterogeneities, albeit of another type, also control the time scales of the jamming transition of a horizontally vibrated granular monolayer~\cite{lechenault2008csa}. However these experiments don't provide any direct measurement of a mechanical response function. Conversely many experiments report on the stress-strain relation in dense granular packings~\cite{howell1999sfg}, but don't have an easy access to the dynamics of individual grains. Investigating the drag of an intruder, the subject of this paper, is a possible way of bridging the gap between dynamical and mechanical properties. Experiments in colloids~\cite{habdas2004fmp}, foams~\cite{dollet2005tdf} and granular media~\cite{zik1992msv,albert1999sap,geng2005sdt,dalton2005ssf} as well as simulations of structural glasses~\cite{hastings2003dfg} were performed along this line. For loose packings and large drag, that is in the so-called fluidized regime, the velocity dependence of the drag force $F$ follows Stokes' law, $F\propto V$~\cite{zik1992msv}. For denser packings, experiments report either $F\simeq cst$ or $F\propto \ln(V)$ ~\cite{albert1999sap,geng2005sdt,dollet2005tdf,dalton2005ssf} when the velocity is fixed, and $F=F_Y+V^{\gamma}$, with $\gamma\le 1$ and $F_Y$ a finite yield force, when the force is imposed~\cite{habdas2004fmp,hastings2003dfg}. In both cases, the dynamics has been described as very intermittent. Stress fluctuations have been investigated in details in~\cite{geng2005sdt}, but very little is known about the displacement field and the velocity fluctuations.

\begin{figure}[b] 
\center
\vspace{-0.5cm}
\includegraphics[width=0.51\columnwidth]{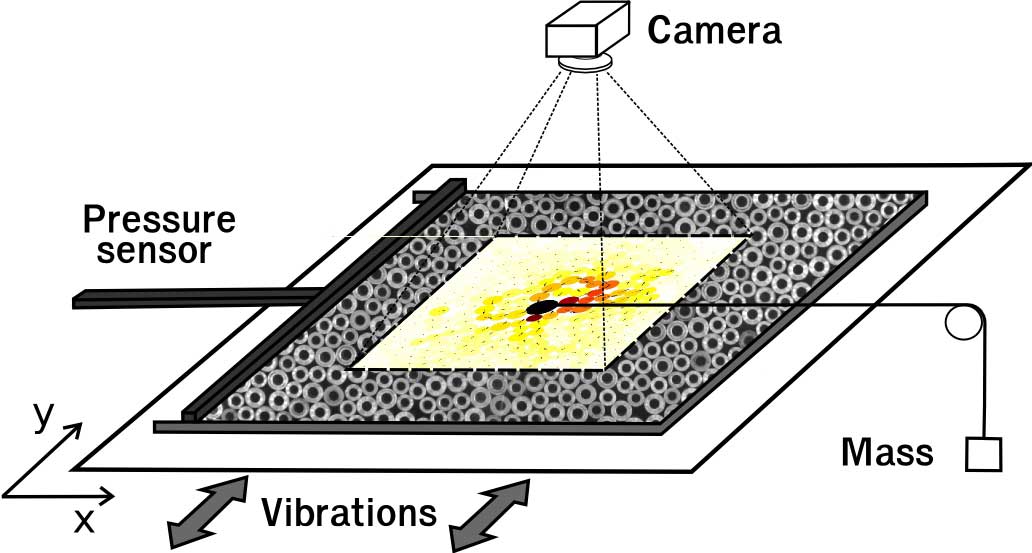}
\includegraphics[width=0.40\columnwidth]{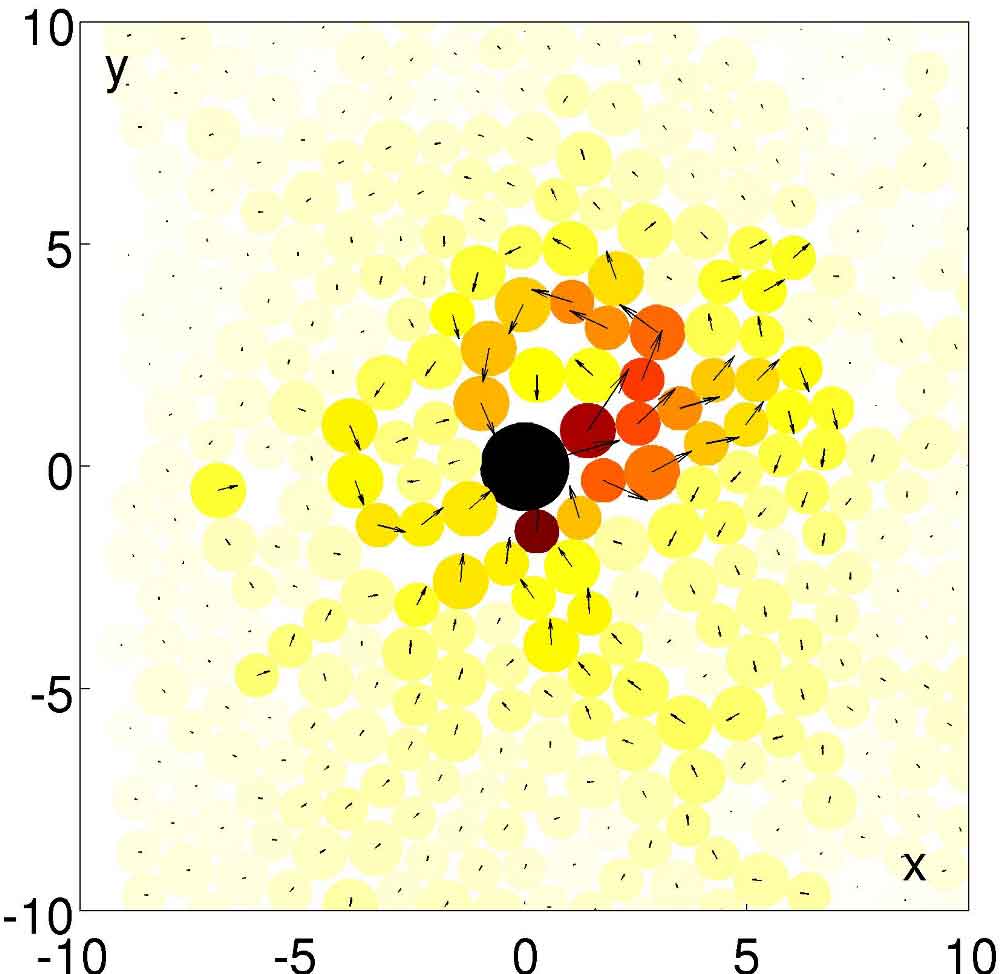}
\caption{\textbf{Left}: a monolayer of disks is vibrated horizontally, while dragging an intruder at constant force. Both the intruders and the surrounding grains are tracked by a CCD camera; \textbf{Right}: a strongly intermittent and heterogeneous response is observed, see text for details}
\label{fig:setup}
\vspace{-0.5cm}
\end{figure}

In this letter we study the dynamics of an intruder dragged by a constant force, in a bidisperse mono-layer of horizontally vibrated grains. The experiment is run in the same set up (fig.~\ref{fig:setup}) and following the same protocol as in~\cite{lechenault2008csa}, where the jamming transition has been identified without ambiguity: at $\phi_J$ the pressure measured in the absence of vibration vanishes and dynamical heterogeneities exhibit a critical behaviour. Here, we observe that close to $\phi_J$, the intruder motion is strongly intermittent, with widely fluctuating velocities. We analyze these fluctuations which resemble very much a "crackling noise" signal~\cite{sethna2001cn} and show that the jamming transition is signed by critical fluctuations of the intruder motion. Investigating the displacements and the free volume fields around the intruder, we conclude that the transition separates a regime dominated by local free volume rearrangements from a regime dominated by the rearrangements of the force network. We also show that the above transition is distinct from fluidization~\cite{dalton2005ssf}, observed at a looser packing fraction which depends on the applied force, when the intruder recovers a continuous motion. 

% -- FIGURE 0001 ------------------------------

The experimental setup (fig.~\ref{fig:setup}) has been described elsewhere and we shall only recall here its most important elements together with the modifications imposed by the drag of the intruder. A monolayer of $8500$ bi-disperse brass cylinders of diameters $d_{small} = 4/5 d_{big} = 4\pm0.01{\rm mm}$ lays out on a horizontal glass plate vibrated  horizontally ($f=10{\rm Hz}$, $A=10{\rm mm}$). The grains are confined in a cell fixed in the laboratory frame. The packing fraction, $\phi$, can be varied by tiny amounts ($\delta\phi/\phi \sim 5.10^{-4}$) and the pressure exerted on the moving lateral wall is measured by a force sensor. The intruder consists in a larger particle of same height ($d_{intruder} = 2.d_{small}$) introduced in the system and pulled by a mass via a pulley perpendicularly to the vibration. The fishing wire, which stands over the other grains, doesn't disturb the dynamics.
Most of the results presented here are related to experiments performed at a constant force ($F_1=0.67N, F_2=1.48N, F_3=2.62N$) and varying the packing fraction, but we also conducted experiments at constant packing fractions, ($\phi_1=0.8383, \phi_2=0.8394, \phi_3=0.8399$) increasing the force from $0.67 N$ to $14.25N$. For comparaison, the total weight of the grains is $23.11 N$ and the force registered at the wall when the grains are highly compressed is of the order of $50 N$. The time unit is set to one plate oscillation while the length unit is chosen to be the diameter of the small particles.

% -- FIGURE 0002 ------------------------------
\begin{figure}[t] 
\center
\begin{minipage}{0.49\columnwidth}
\includegraphics[width=\textwidth, height=1.05\textwidth]{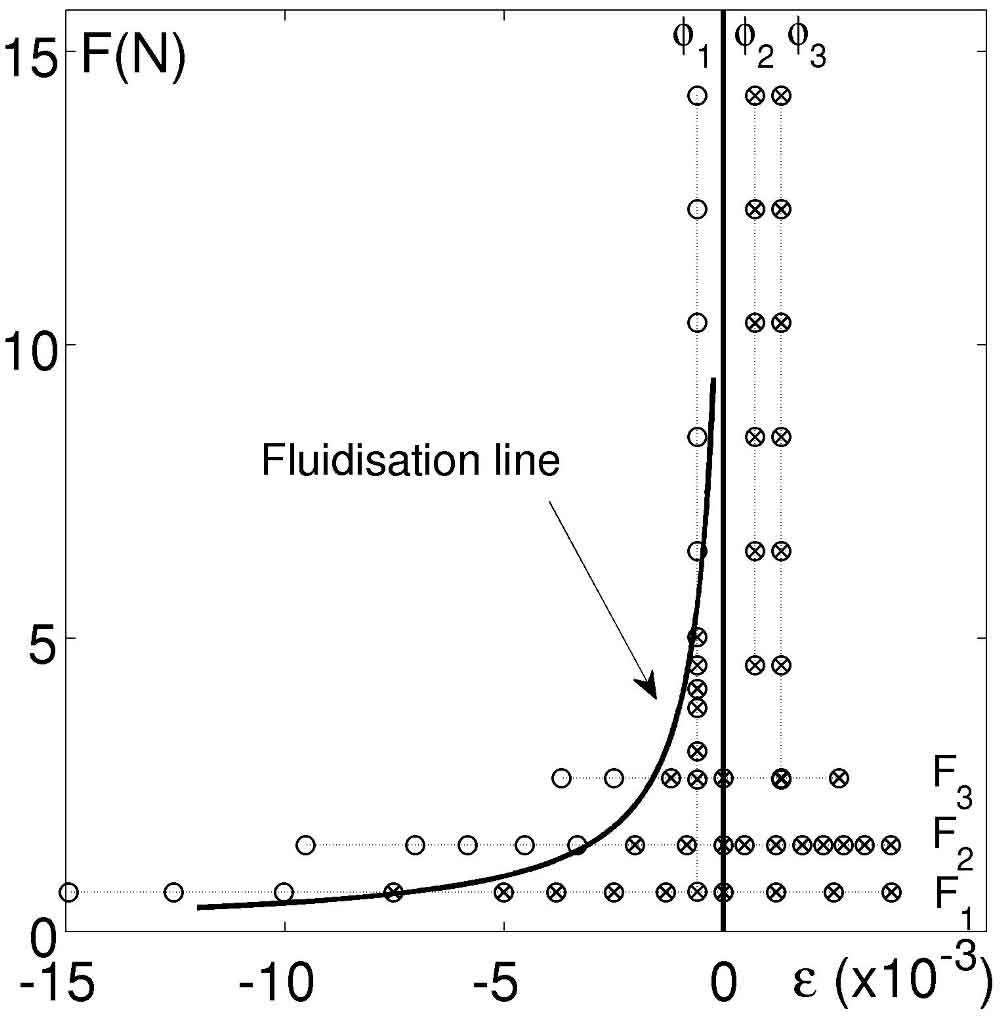}
\end{minipage}
\begin{minipage}{0.49\columnwidth}
\includegraphics[width=\textwidth, height=\textwidth]{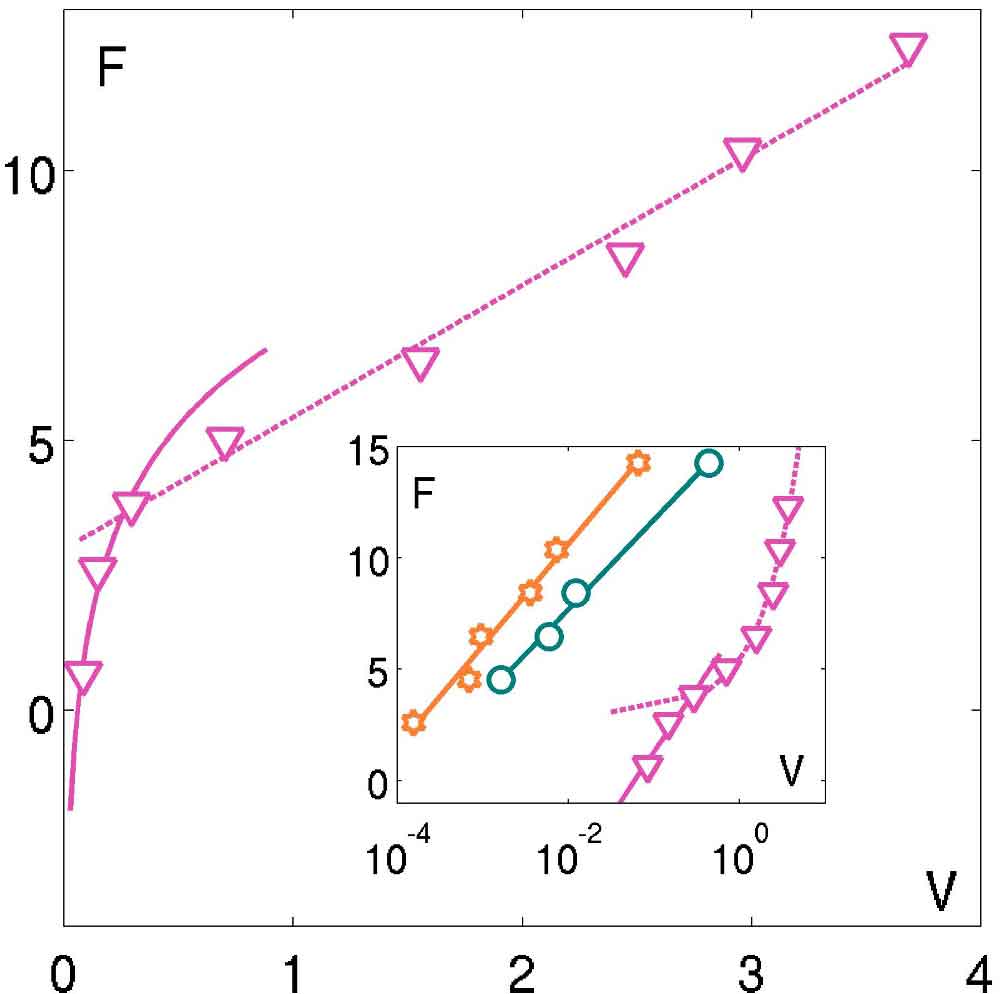}
\end{minipage}
\caption{From flow to jammed states: \textbf{Left} Parameter space, (force $F$ - relative packing fraction $\epsilon=(\phi-\phi_J)/\phi_J$): at low $\phi$ and large $F$,($\circ$) the packing is fluidized, the intruder motion is continuous and $F\propto V$; at large $\phi$ and low $F$ ($\otimes$), the intruder exhibits an intermittent motion and $F\propto \ln V$. The horizontal and vertical doted lines indicate the path followed in the parameter space in the present study. \textbf{Right} $F$ versus $V$ along the paths $\phi_1 (\medtriangledown)$ , $\phi_2(\circ)$, $\phi_3(\medstar)$; \textit{Inset}: same in log-lin.}
\label{fig:param}
\vspace{-0.1cm}
\end{figure}

Starting from a low packing fraction $\phi$, we gradually compress the system until it reaches a highly jammed state following the same protocol as in~\cite{lechenault2008csa}. Then we stepwise decrease the volume fraction. In the absence of intruder, it was shown that the average relaxation time increases monotonically with the packing fraction, while the dynamics exhibits strong dynamical heterogeneities, the length-scale and time-scale of which exhibit a sharp peak at an intermediate packing fraction. The pressure measured at the wall in the absence of vibration falls to zero precisely below that packing fraction, hence called the jamming transition $\phi_J$. In the present study, the intruder is inserted at its initial position in place of one big and two small grains before each downward step in packing fraction and the system is kept under vibration until the pressure has recovered its value in the absence of the intruder. Only then the force is applied on the wire and the intruder is dragged through the cell, while its motion together with that of a set of $1800$ surrounding grains in the center of the sample is tracked by a digital video camera triggered in phase with the oscillations of the plate. 

% -- FIGURE 0003 ------------------------------
\begin{figure}[t] 
\center
\includegraphics[width=0.49\columnwidth, height=0.49\columnwidth]{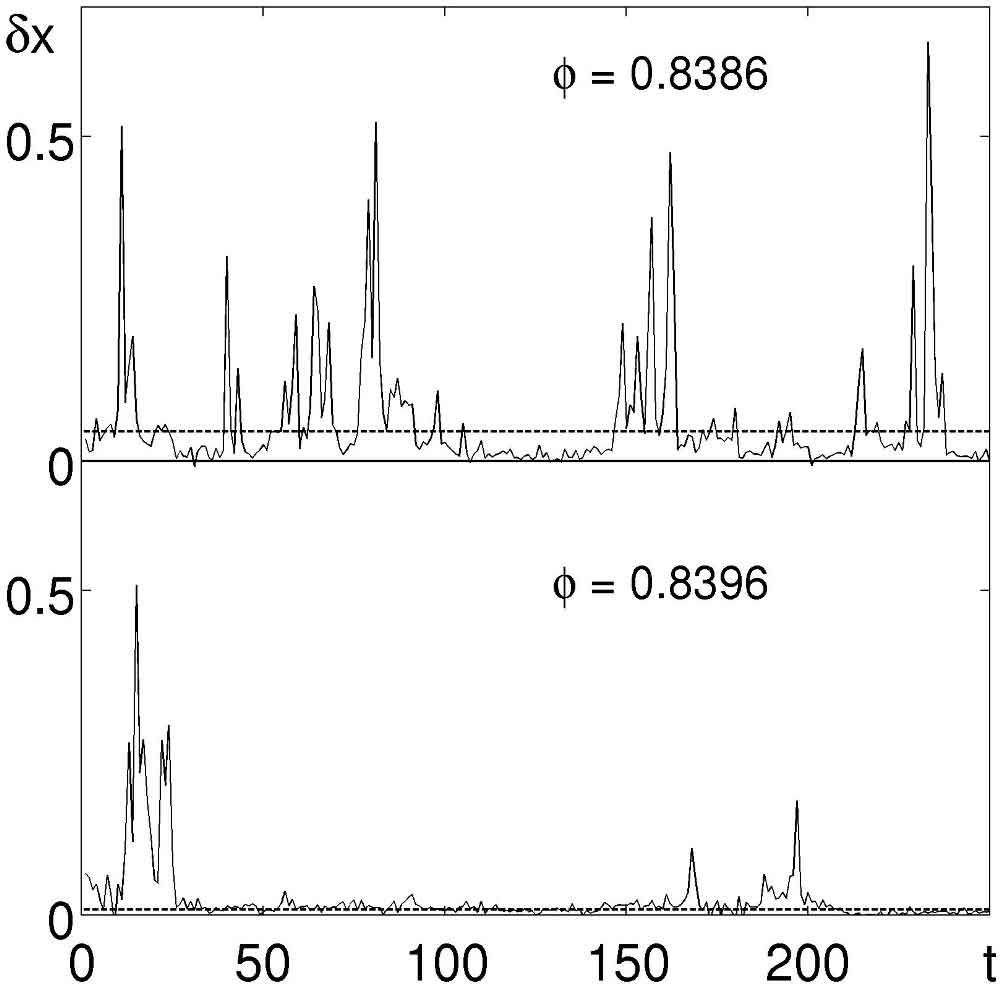}
\includegraphics[width=0.49\columnwidth, height=0.49\columnwidth]{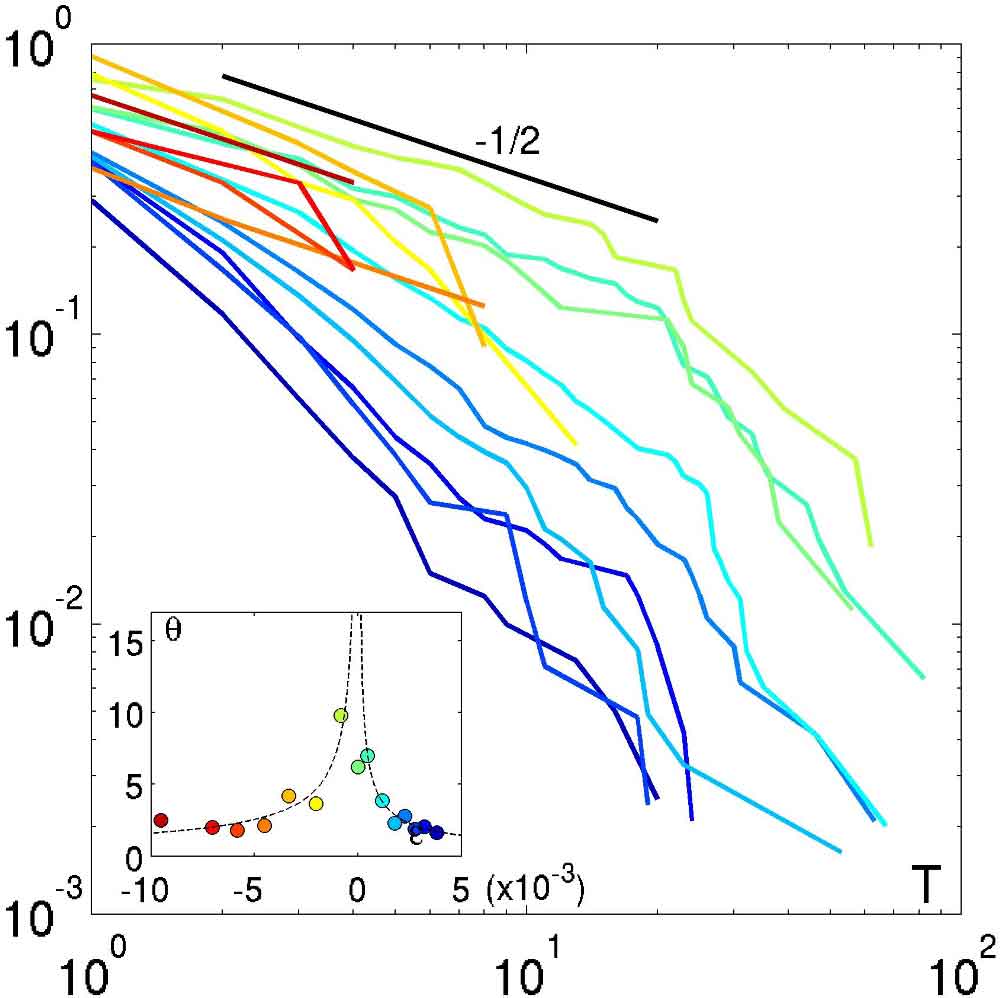}
\includegraphics[width=0.49\columnwidth, height=0.49\columnwidth]{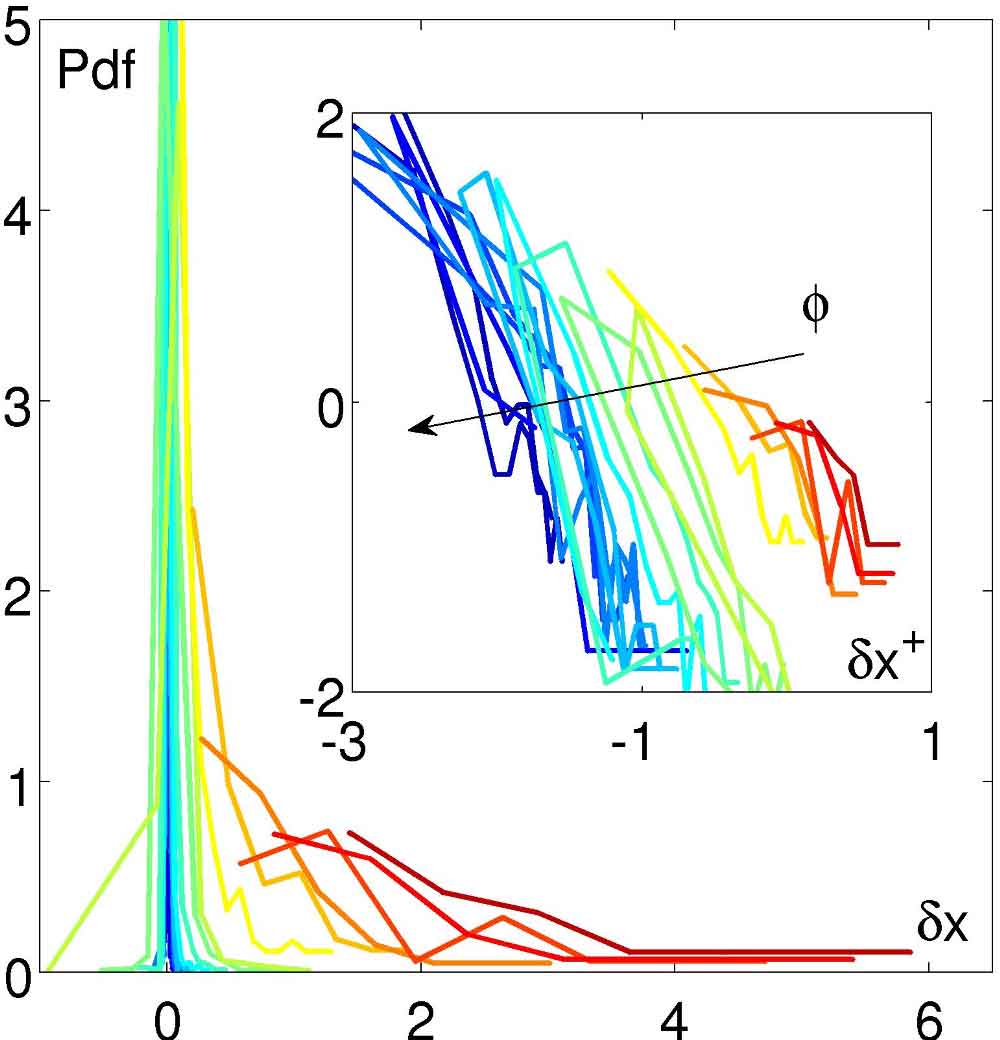}
\includegraphics[width=0.49\columnwidth, height=0.49\columnwidth]{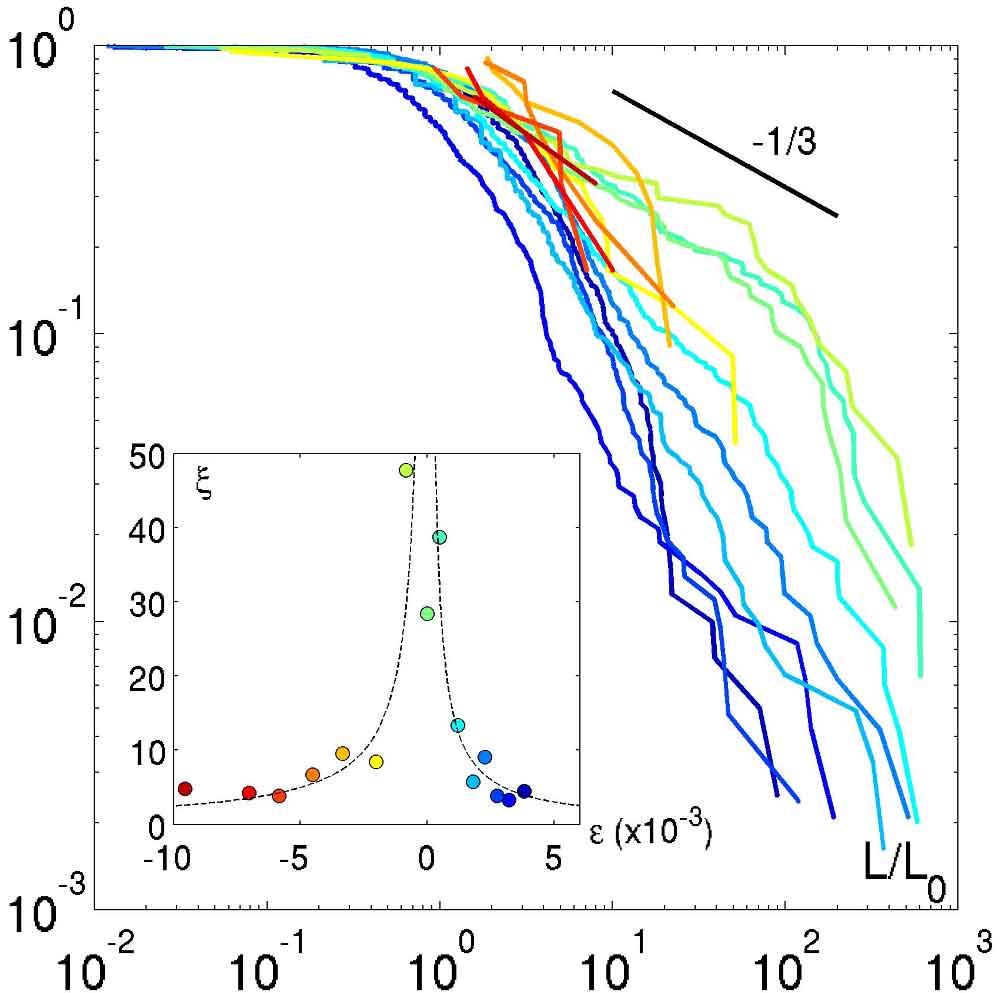}
\includegraphics[width=0.49\columnwidth, height=0.49\columnwidth]{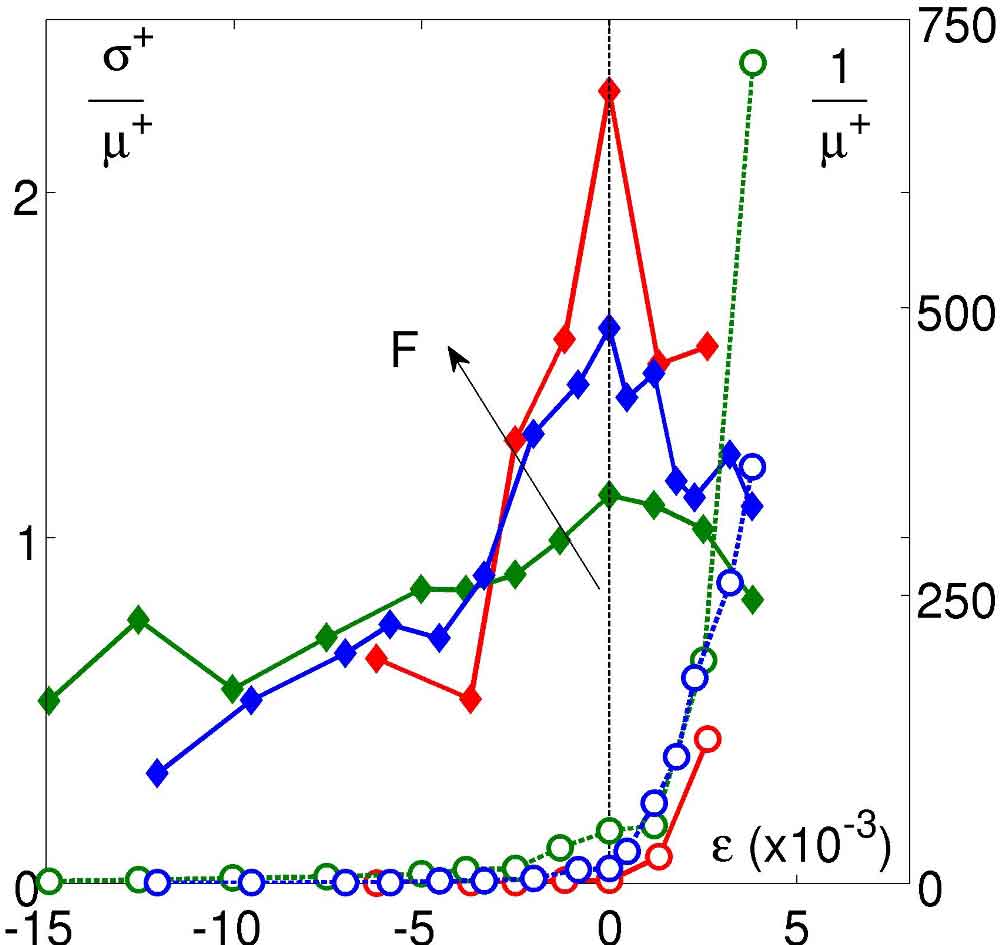}
\includegraphics[width=0.49\columnwidth, height=0.49\columnwidth]{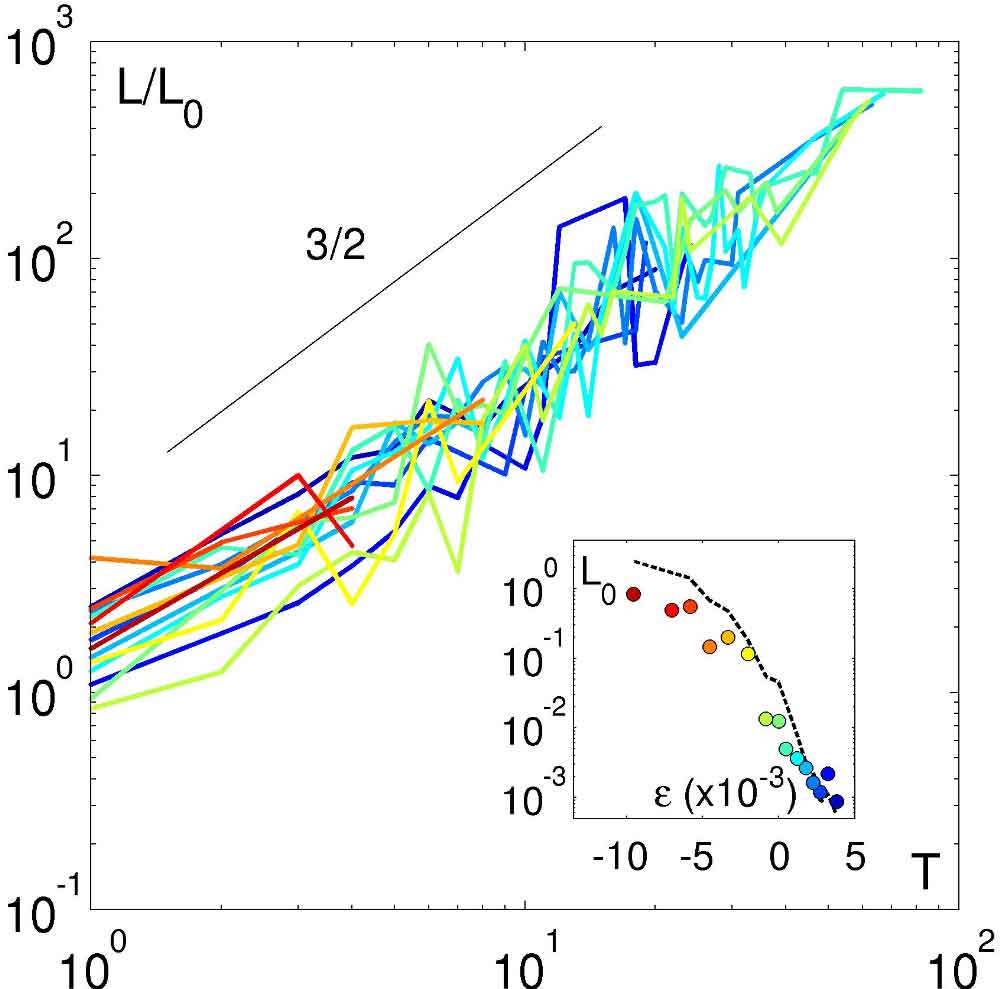}
\caption{
\textbf{Top-Left} Intruder instantaneous displacements $\delta_x(t)$ for two different packing fractions.
\textbf{Middle-Left} Distributions of $\delta_x$ for $15$ packing fractions. \textit{Inset}: same in log-log. 
\textbf{Bottom-Left} Inverse average $1/\mu_+$ ($\circ$ - right axis) and standard deviation over average $\sigma_+/\mu_+$ ($\filleddiamond$ - left axis) of $\delta_x^+$ as a function of the reduced packing fraction $\epsilon$ for the three applied force $F_1$ (red), $F_2$ (blue) and $F_3$ (green). 
\textbf{Top-Right} Cumulated distributions of the durations $T$ of the displacement bursts for different packing fraction ; \textit{Inset}: Cutoff of the distributions $\theta(\epsilon)$ versus $\phi$.
\textbf{Middle-Right} Cumulated distributions of the rescaled size $L/L_0$ of the bursts for different packing fraction; \textit{Inset}: Cutoff of the distributions $\xi(\epsilon)$ versus $\phi$. 
\textbf{Bottom-Right} Scaling of the size $L$ of the bursts with their duration $T$. The avalanche sizes are rescaled by a factor $L_0$. \textit{Inset}: $L_0$ and $V$ (dark dotted line) follow the same evolution with $\phi$. (In all case, $F=F_2$ when not specified.)}
\label{fig:intruder}
\vspace{-0.1cm}
\end{figure}

At low packing fraction, and large enough force, the intruder motion is continuous. When increasing the packing fraction, it becomes intermittent above some threshold, which increases with the applied force (fig~\ref{fig:param}-a). Performing experiments at a given packing fraction and increasing the force, one observes that the applied force is proportional to the average velocity of the intruder $F\propto V$ in the continuous motion regime, whereas $F\propto \ln V$ in the intermittent one (fig~\ref{fig:param}-b). We thereby identify this transition with the fluidization one~\cite{dalton2005ssf}. Note that for the largest packing fraction $\phi_2$ and $\phi_3$, we could not observe the fluidization. As a result, there is a very strong contrast between the continuous motion of the intruder observed at $\phi=\phi_1$ for large enough forces, say $F=10N$, and the strongly intermittent one observed at $\phi=\phi_2$ for the same large force. Typically the intruder averaged velocity looses three orders of magnitude and its velocity flucutations gain more than five orders of magnitudes, while $\phi$ is increased by less than $1\%$, already suggesting the existence of a sharp transition. 

We now focus on the experiments performed at constant and rather small force. When looking at the intruder displacements $\delta_x$ along the dragging direction during one vibration cycle (fig.~\ref{fig:intruder}-top-left), one immediately notices very strong fluctuations, with bursts of widely fluctuating magnitude. More quantitatively, the probability density functions of $\delta_x$ (fig.~\ref{fig:intruder}-middle-left) exhibits an important skewness towards the positive displacements. We characterize the positive part of the distribution \ie the displacements in the direction of the drag force, $\delta x^+$, by computing the average value $\mu^+=\left\langle \delta x^+\right\rangle $ and the relative fluctuations $\sigma^+/\mu^+=\left\langle (\delta x^+ - \mu^+)^2\right\rangle ^{1/2}/\mu^+$. One observes (fig.~\ref{fig:intruder}-bottom-left) that $1/\mu^+$ increases continuously by three orders of magnitude, while varying the packing fraction of only a few percent, $\delta \phi/\phi = 2.10^{-2}$, and that $\sigma^+/\mu^+$ exhibits a peak at an intermediate packing fraction. Both behaviours are directly reminiscent of what has been recalled above for the dynamics in the absence of intruder. We could check indeed that the peak observed in the fluctuations of the intruder motion coincides with a vanishing pressure in the absence of vibration and thereby locate it at $\phi_J$ without ambiguity. Note however that the precise value of $\phi_J$ depends on the precise packing that has been selected when the system has been compressed and that the small and non monotonous variations of $\phi_J=0.8369, 0.8386, 0.8379$ for the three forces $F_1, F_2, F_3$ together with the small difference with the value $\phi_J=0.842$ reported in~\cite{lechenault2008csa} must be attributed to differences in the initial conditions and more generally in the compression protocol (see~\cite{zamponi2008mft,berthier2008cnh} for a detailed discussion in the case of hard spheres). Figure~\ref{fig:intruder}-bottom-right displays both $1/\mu^+$ and $\sigma^+/\mu^+$ as a function of $\epsilon=(\phi-\phi_J)/\phi$ the relative distance to $\phi_J$ for the three dragging forces. The peak in the fluctuations is clearly separated from the divergence of $1/\mu^+$ and is a new signature of the jamming transition independent from the one provided by the study of the dynamical heterogeneities. 

% -- FIGURE 0004 ------------------------------
\begin{figure}[t] 
\center
\includegraphics[width=0.49\columnwidth, height=0.49\columnwidth]{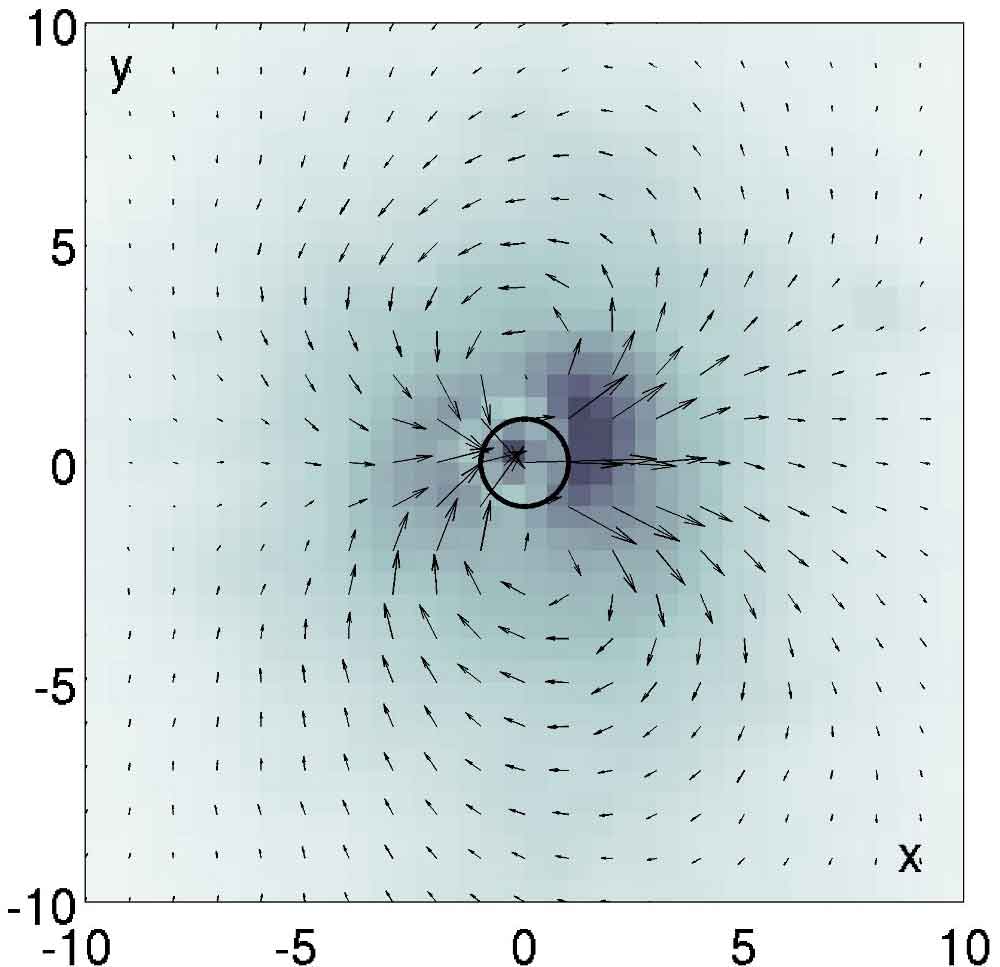}
\includegraphics[width=0.49\columnwidth, height=0.49\columnwidth]{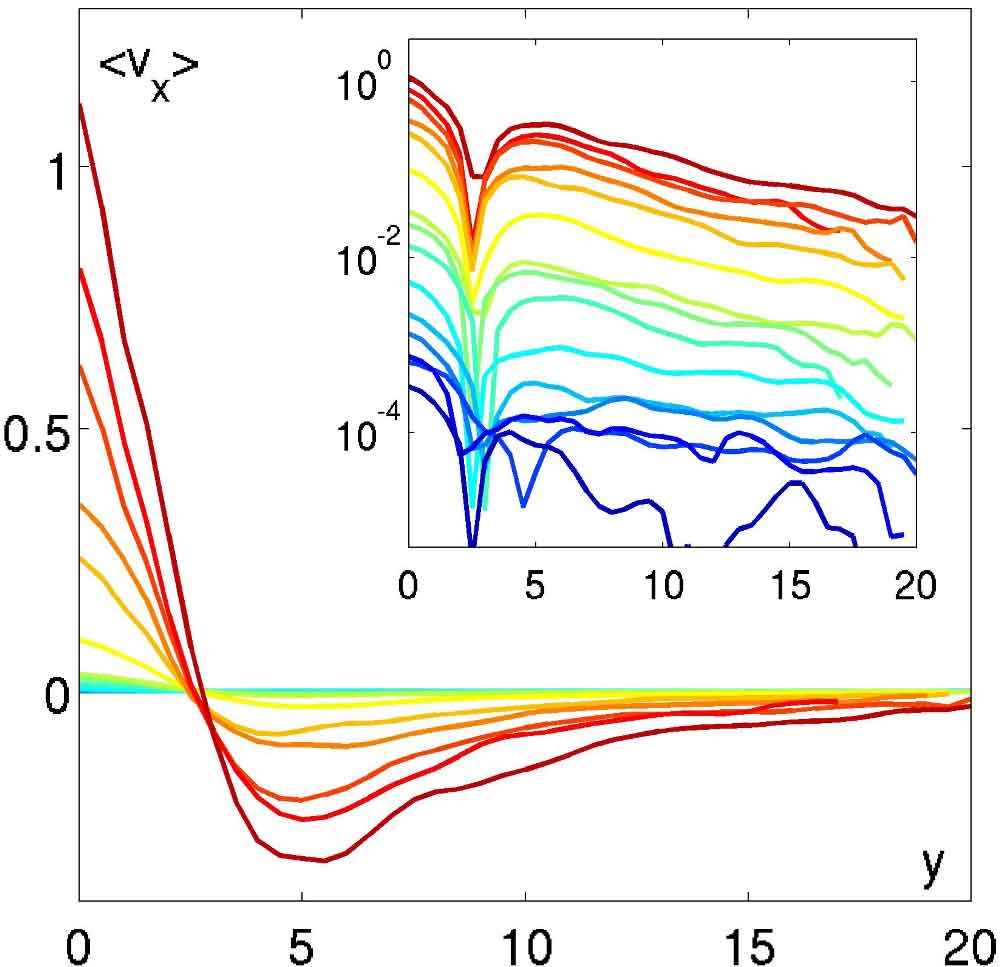}
\includegraphics[width=0.49\columnwidth, height=0.49\columnwidth]{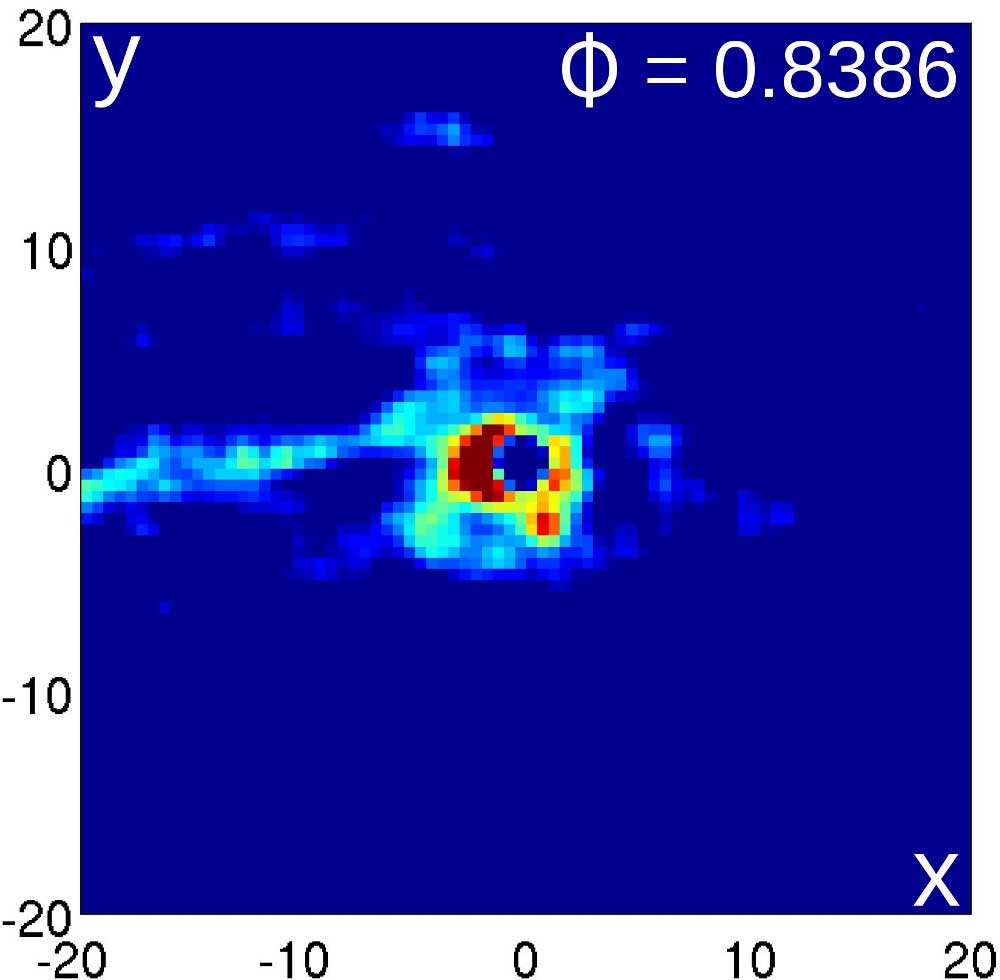}
\includegraphics[width=0.49\columnwidth, height=0.49\columnwidth]{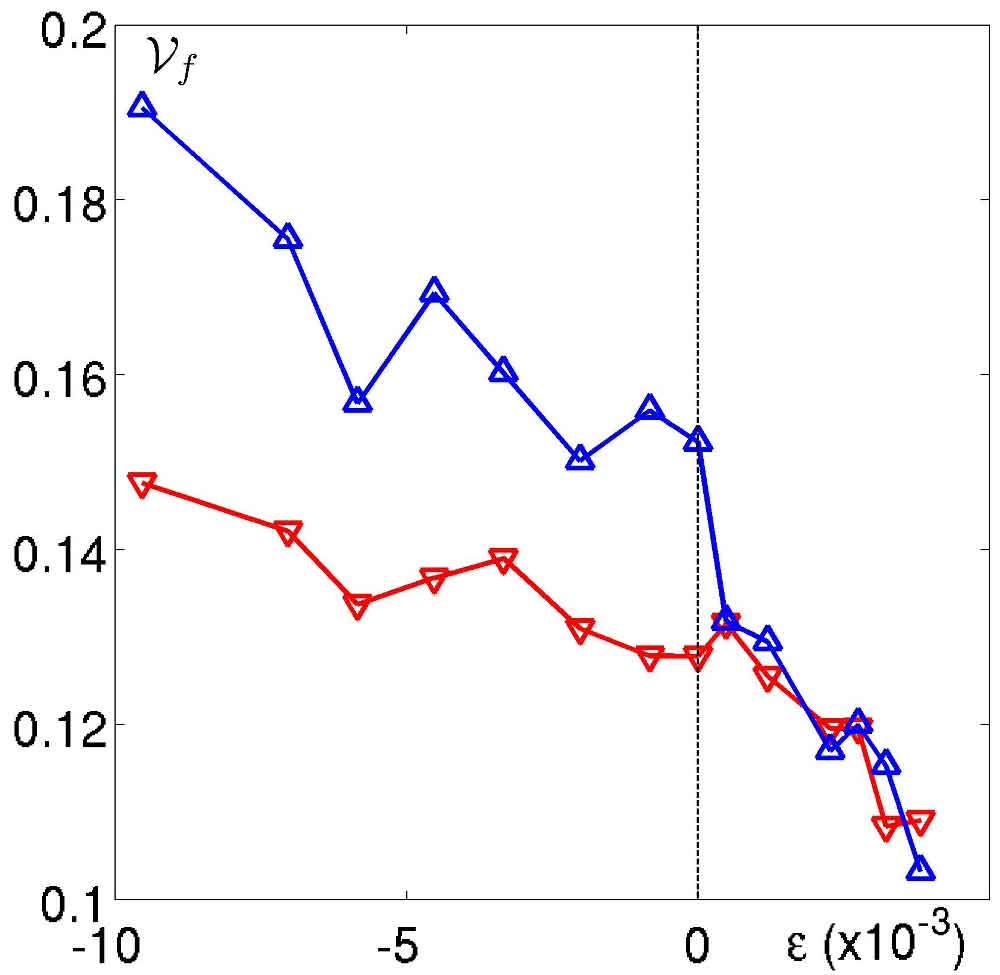}
\caption{Displacements and free volume around the intruder (the intruders goes from left to right; $F=F_2$).
\textbf{Top-Left} Interpolated average displacement field : the darker, the faster. 
\textbf{Top-Right} y-profiles of the average velocity along the drag direction for different packing fractions. 
\textbf{Bottom-Left} Average free volume field at $\phi=\phi_J$.
\textbf{Bottom-Right} Average free volume vs $\epsilon$ in front (red $\medtriangledown$) and behind (blue $\medtriangleup$) the intruder.}
\label{fig:field}
\vspace{-0.1cm}
\end{figure}

To complete the characterization of the intruder motion, we analyse the duration $T$ and the size $L$ associated with the displacement bursts, as is commonly done in crackling noise or Barkhausen noise experiments~\cite{sethna2001cn}.
We impose a given reference level $\delta_x^0$ and define bursts as zones where
$\delta_x$ is above this level. The duration $T$ of a given burst is defined
as the interval within two successive intersections of $\delta_x$
with $\delta_x^0$, while the size $L$ is defined as the integral of $\delta_x$
between the same points. Despite some weakness of our statistics, the cumulated distributions of $T$ (fig~\ref{fig:intruder}-top-right), respectively of $L/L_0$ (fig~\ref{fig:intruder}-middle-right), where $L_0(\phi)$ is a scaling factor following the same evolution as $V(\phi)$, can be described as power laws truncated by a scaling function : $T^{-\alpha}f(T/\theta(\phi))$, resp. $(L/L_0)^{-\beta}g(L/\xi(\phi))$. The cut-off $\theta(\phi)$, resp. $\xi(\phi)$, estimated by computing the experimental averages of $T$, resp. $L/L_0$ display a diverging behaviour at $\phi_J$: $\theta(\phi) \sim \lvert\phi-\phi_J\rvert^{-\eta}$, resp. $\xi(\phi)\sim \lvert\phi-\phi_J\rvert^{-\nu}$. Our statistics are not large enough to extract precisely the exponents $\alpha$, $\beta$, $\eta$ and $\nu$; however the dynamical exponents $z$, defined by $L/L_0(\phi)\sim T^{1/z}$ (fig~\ref{fig:intruder}-top-bottom) is reasonably well determined, $z=2/3$ and estimates of $\alpha=1/2$, $\beta=1/3$, $\eta=2/3$ and $\nu=1$ are consistent with the data and satisfy the relations $\alpha z = \beta$ and $\eta = \nu z$. Also, the same analysis performed on the kinetic energy of the surrounding grains -- not shown here -- is consistent with the above determination.
 
Finally, we characterize the dynamics around the intruder. The averaged displacement field (fig.~\ref{fig:field}-top-left) is composed of two symmetric recirculation vortices. This pattern is very robust as evidenced by the shape invariance of the displacement profiles along the direction perpendicular to the drag (fig.~\ref{fig:field}-top-right). In particular, the exponential decay of the displacement amplitude keeps the same characteristic length across the transition (see inset).
Such a smooth and continuous behaviour is in contrast with the existence of the sharp transition described above and must be related to the similar absence of signature of the transition when considering the average relaxation time or $\mu^+(\phi)$. Again the transition is to be found in the strongly heterogeneous instantaneous displacement field (fig.~\ref{fig:setup}-right), which exhibits characteristic chain-like motions. This tendency of forming chain-like motions is strongly enforced for packing fraction larger than $\phi_J$. We also compute the averaged free volume, extracted from the Laguerre's tessellation of the packing, around the intruder (figure~\ref{fig:field}-bottom-left) and observe a small asymmetry between the front and the back of the intruder. Computing for instance the average free volume in a small window in front of and behind the intruder (see figure~\ref{fig:field}-bottom-right), one sees that below $\phi_J$ there is a significant excess of free volume leaving like a "wake" far behind the intruder, whereas above $\phi_J$ this asymmetry rapidly vanishes together with a strong decrease of the available free volume.

Altogether, the following pictures emerge. The jamming transition, which is marked by a critical behaviour of the fluctuations of the intruder displacements separate a regime where the intermittent motion is dominated by rapid reorganization of the free volume from one where the motion takes the form of chain-like structures, which are very much reminiscent of the strong force network suggesting the dominant role of the stress fluctuations~\cite{Howell1999sfgc}. As compared to previous studies of the motion of an intruder in a granular packing~\cite{zik1992msv,albert1999sap,geng2005sdt,dalton2005ssf}, we have confirmed the transition from a linear viscous-like dependence of the dragging force with the velocity in the fluidized regime to a logarithmic dependence in the intermittent one.  The originality of the present study is to have unveil critical features {\it inside} the intermittent regime and associate them with the jamming transition. From a more general point of view, the observation of "crackling noise" statistics suggests a possible deeper correspondence in the underlying physical properties with other intermittent phenomena such as the creep motion observed under yield stress in amorphous media~\cite{falk2004stz}, the subcritical material failure~\cite{bonamy2008cdmf} and more generally, the pinning-depinning transition~\cite{chauve2000cad}. Further investigations in the case where the intruder is dragged at constant velocity through a spring, should provide new insight in this matter. We believe that performing the kind of micro-rheology experiment sketched out in the present study is a promising path for a better understanding of the mechanisms at play at the jamming transition of frictional systems.

We would like to thank L. Ponson for having suggested the "crackling noise" analysis, E. Bouchaud, and S. Aumaitre for helpful discussions. We also thank V. Padilla and C. Gasquet for technical assistance on the experiment. This work was supported by ANR DYNHET 07-BLAN-0157-01. 

\bibliography{biblio_glass}

% \begin{thebibliography}{0}
% \end{thebibliography}

\end{document}